# Comments on theoretical foundation of "EM Drive"

C.-W. Wu

Institute of Mechanics, Chinese Academy of Sciences, No.15 BeisihuanXi Road, Beijing 100190, China

Tel: 86-10-8254-4271 Fax: 86-10-6256-1284

E-mail: chenwuwu@imech.ac.cn & c.w.wu@outlook.com

**Abstract**
The concept of EM Drive has attracted much attention and groups of work have been conducted to prove or verify it, of which the published experimental outcome is criticized in great details while the theoretical foundation has not been discussed. The present essay investigates on the theoretical derivations of the net thrust in the "EM drive" and reveals the self-contradiction arising at the very start, when the law of conservation of momentum was utilized and opposed simultaneously.

The "EM Drive" herein discussed refers to the "radio frequency (RF) resonant cavity thruster" proposed by the British engineer Roger Shawyer. This concept has attracted a lot of attention [1] and some experimental outcome [2] have been claimed to be supportive, although much of the details have been questioned [3]. The present author would like to discuss the paradox in the physical foundation of the net thrust in the "EM drive".

Basically, the theoretical derivations by Shawyer [4] and Yang [5] are inaccurate, of which the typical concept could be sketched as Fig. 1.

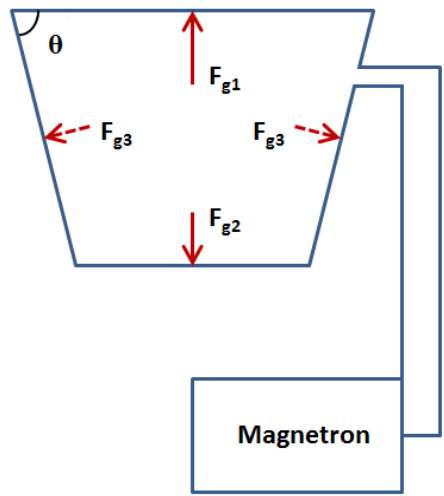

**Figure 1 Theoretical sketch of the EM drive [4, 5]**

Shawyer defines the net thrust as [4]
$$F_{thrust} = F_{g1} - F_{g2}. \tag{1}$$
Wherein, the force $F_{g3}$ acted upon the side wall of the so-called waveguide has been ignored without any explanation.

In comparison, Yang et al assumes that $F_{g3}/F_{g1}$ is slight enough to be negligible and therefore [5]
$$F_{thrust} \approx F_{g1} - F_{g2}. \tag{2}$$

Hence after, Yang et al [6] updated the formula to include the side force $F_{g3}$ and write the resultant "electrical-field force" as

$$F_e = (\iint \tfrac{1}{2}\varepsilon_0 E^2 ds)_{S_1-S_2} + (\iint \tfrac{1}{2}\varepsilon_0 E^2 cos\theta ds)_{S_3}. \qquad (3)$$

As well as the resultant "magnetic-field force" as

$$F_m = (\iint \tfrac{1}{2}\mu_0 H^2 ds)_{S_1-S_2} + (\iint \tfrac{1}{2}\mu_0 H^2 cos\theta ds)_{S_3}. \qquad (4)$$

It is noteworthy that the so-called electrical-field force and magnetic-field force described by Yang [6] are rather strange in physics as no electric charge is involved, which might have been confounded with the radiation pressure that resulted from the momentum change rate of incident photons. Moreover, the dimension of the right hand side of (3) or (4) differs from that of force in absence of some velocity as denominator, which should be the light velocity *c*. One can further see through (3) and (4) that, the sum of the so-called electrical-field force and magnetic-field force obviously amounts to the surface integral of the Poynting vector that represents power density of the electromagnetic wave. That is to say, the participation of so-called electromagnetic resonant does not change the physical essence of the "electromagnetic field force" as radiation pressure, which is exactly one phenomenon of conservation of momentum.

Sure enough, the first term in the electrical-field force or magnetic-field force should be of almost identical magnitude but inverse sign with that of its second term. Herein, the word "almost" is utilized to respect the preciseness of science as there might be a very slight lag in time (of magnitude $10^{-8}$ s) considering the successive incidences of every photon upon the top, bottom or side wall. Macroscopically, such small lag in time of any single photon would never produce net the dimension on the right hand side of (3) or (4) differs from that of force in absence of a velocity thrust because the analogous actions of the vast disorder photons would make such difference random and unobservable.

In other words, the magnitude of the component of $F_{g3}$ along the axial direction ($F_{g1}$ and $F_{g2}$) would always be

$$F_{g3} \times cos\theta = |F_{g1} - F_{g2}|. \qquad (5)$$

And, the apparent resultant force acted on the cavity wall should be zero.

$$\boldsymbol{F}_{thrust} = \boldsymbol{F}_{g1} + \boldsymbol{F}_{g2} + \boldsymbol{F}_{g3} = 0. \qquad (6)$$

For a closed cavity as considered by Shawyer [4] and Yang [5, 6]. It should also be the case for any similar closed system with some interior stable electromagnetic field pattern developed by multiple absorptions, reflections and interferences of the electromagnetic waves if the math work is done accurately. Because the wall pressure exerted by electromagnetic field is actually the radiation pressure resulted from the change in momentums of the incident photons upon the wall [7]. And the change in momentums of photons is directly determined by the interaction between the incident photons and the system itself, including the absorption, reflection and emission of photons, which should definitely obey conservation of momentum. As we know, the total momentum of the whole closed system should not be altered only by the internal interactions.

In a sum, the disregard of the equilibrium relationship as indicated by (5) and (6) might be the very root why the previous works on EM drive have not made any physical sense. Again, we should never count on that some phenomenon of conservation of momentum would in turn violate the law and give us a big surprise.

**Acknowledgements**

The reviewers and editors are appreciated for the great advices in revision. Thanks are also due to my colleagues, Prof. Huang, Dr. Liu and Mr. Liu from Chinese Academy of Sciences who asked me to pay attention to the event of "EM drive".